\documentclass[reprint,aps]{revtex4}
\usepackage{amssymb,epstopdf,amsmath,enumerate}   
\newcommand{\Array}[2]{\left(\begin{array}{#1}#2\end{array}\right)}

\begin{document}

\title{Mass Matrix Rules and the Flat Pattern of Quarks}
    \author{Ying Zhang\footnote{E-mail:hepzhy@mail.xjtu.edu.cn.}}
    \address{School of Science, Xi'an Jiaotong University, Xi'an, 710049, China}

\begin{abstract}
Seeking mass patterns is a key to decoding the unknown flavor puzzles in particle physics.
Inspired by quark hierarchical masses, the mass matrix can universally be factorized into a family-diagonal phase matrix $K_L^q$ and a real symmetric matrix $M_N^q$ characterized by only two parameters. 
The factorized structure provides model-independent rules to the mass matrix. 
We demonstrate that the large $\delta_{CP}$ naturally arises from the degeneracy of the first two quark families in the mass hierarchy limit. As an application, the flat pattern with elements close to unity in the matrix is checked by fitting quark masses and the CKM mixing. It achieves a precise description of flavor structure with minimal parameters.



\end{abstract}

\maketitle

\section{Motivation}
\label{sec.intro}
Although the Standard Model (SM) of particle physics has successfully described the gauge interactions of strong and electroweak, it has not been providing a clear understanding of the Yukawa interactions. 
The SM relies on complex, family-dependent Yukawa couplings to explain fermion masses and mixing, introducing numerous redundant parameters without a clear theoretical basis or pattern.
It is believed that an unknown mechanism controlling the mixing and mass hierarchy may be hidden in mass matrixes \cite{FroggattNPB1979,FeruglioEPJC2015,ZupanCERN2019}.

Many models have been proposed to address these flavor puzzles \cite{CrivellinPRD2013,AlonsoJHEP2018,BlechmanJEHP2010,AntonelliPR2010}. 
Some are motivated by discrete and/or continuous symmetries in the flavor mixing \cite{SFKingRPP2013,HarrisonPLB2002,WangPRD2021}, while others focus on hierarchical masses \cite{RandallPRL1999,WeinbergPRD2020,MohantaJHEP2024}.
With increasingly precise measurement of masses, mixing angles, and CP violation, there is almost no parameter space left for these flavor models.
A popular way to evade precise flavor constraints is to introduce more parameters within
three-family framework of the SM or in an extended flavor space.
In the quark sector, flavor observables include six masses, three mixing angles, and one CP-violating phase.  In the lepton sector with extended Dirac neutrino, the same number of observables exists.
Models with more than ten free parameters inevitably predict new physics beyond the SM, yet these predictions must face the challenge of null results from experimental searches. On the other hand, models with fewer parameters may impose constraints on masses and the mixing that are not supported by current phenomenology.
In balancing the fit precision and the number of free parameters, 
no candidate has emerged as superior. They are either ruled out by precise flavor data or introduce too many free parameters beyond what is phenomenologically required. Therefore, it is natural to seek a mass matrix without redundant parameters and to uncover rules and limitations of mass patterns from these observables.

Another challenge lies in explaining large CP violation. $\delta_{CP}$ has a large value about $61^\circ$ in the CKM mixing and about $270^\circ$ in the PMNS mixing \cite{PDG2024}. Because of hierarchical masses, the observed large $\delta_{CP}$ cannot be attributed to small perturbations in a hierarchical framework. It requires CP violation to originate from a degenerate mass spectrum of the first two families.
This serves as an independent checkpoint for flavor structure.

In this paper, we will not adopt the top-down approach inspired by extending the SM from symmetries or some theoretical motivations. Instead, we will use a bottom-up approach to explore the requirements based on observed masses and the mixing.
Although flavor structures exist in both the quark and lepton sectors, we will focus mainly on quarks to avoid the complexities arising from the unknown properties of neutrinos.
We attempt to decode quark patterns from the common characteristic of the mass hierarchy in Sec.\ref{sec.MassPattern}. A set of model-independent rules for the mass matrix is derived, which can apply to all flavor models when heavier new physics particles are integrated out or when the extended flavor space is diagonalized into a block-diagonal form.
In Sec.\ref{sec.correction}, we analyze the flat pattern, characterized by elements close to unity, and systematically incorporate hierarchical corrections up to $\mathcal{O}(h^2)$ to account for the light quark masses. We will illustrate that large CP violation can arise from degenerate masses of the first two families.
A globe fit will be presented in Sec.\ref{sec.fit}, separating hierarchy corrections from the total mixing. 
A summary will be given in Sec.\ref{sec.summary}.

\section{Mass Pattern in the Mass Hierarchy Limit}
\label{sec.MassPattern}
\subsection{Unitarity condition}
After electroweak symmetry breaking, the quark mass matrix $M^q$ (for $q=u,d$) is generated by family-dependent Yukawa couplings. 
$M^q$  can be diagonalized into real eigenvalues by bi-unitary transformations $U_L^q$ and $U_R^q$ as:
\begin{eqnarray}
 U_L^qM^q(U_R^q)^\dag=\text{ diag}(m_1^q,m_2^q,m_3^q). 
 \label{eq.UMU0001}
\end{eqnarray}
The left-handed $U_L^q$ also determines the quark flavor mixing matrix $U_{CKM}$ in charged current weak interaction
\begin{eqnarray}
U_{CKM}=U_L^u(U_L^d)^\dag. 
\end{eqnarray}
However, the right-handed transformation $U_R^q$ is unphysical. For a random unitary matrix $U'$, the mass matrix $M^qU'$ has the same real eigenvalues as $M^q$.
This means that the  $U_R^q$ can be chosen arbitrarily. And $U'$ has no contribution to the mixing $U_{CKM}$. So, the mass matrix $M^q$ cannot be uniquely reconstructed from flavor phenomenology unless additional constraints are imposed to fix $U_R^q$. A convenient choice is to impose the unitarity condition $U_R^q=U_L^q$, which renders $M^q$ a Hermitian matrix, thereby naturally ensuring real eigenvalues.
\subsection{Mass Matrix Reconstruction and Rules}
The physical masses of quarks exhibit a pronounced hierarchy:
	\begin{eqnarray}
	m_1^q\ll m_2^q\ll m_3^q.
	\end{eqnarray}
Defining the mass ratio $h_{ij}^q=m_i^q/m_j^q$ ($i<j$), their values are given by \cite{PDG2024} 
\begin{eqnarray}
h_{12}^u\simeq 0.0017,~ h_{23}^u\simeq 0.0074,~ h_{12}^d\simeq 0.050,~ h_{23}^d\simeq 0.023.
\end{eqnarray}
These hierarchies justify the perturbative expansion of the mass matrix in power of $h_{ij}^q$:
\begin{eqnarray}
	M^q=M_0^q+h_{23}^qM_1^q+h_{12}^qh_{23}^q M_{21}^q+(h_{23}^q)^2M_{22}^q+\mathcal{O}(h^3)
\end{eqnarray}
Here, $M_0^q$ are the leading order mass matrix, and 
$M_1^q,M_{2i}^q$ are corrections of order of $h^q_{ij}$ and $(h^q_{ij})^2$, respectively. 
Note that the leading order effect of $h_{12}^q$ appears at $\mathcal{O}(h^2)$. 

Normalized to the total family mass, Eq. (\ref{eq.UMU0001}) becomes
\begin{eqnarray}
\frac{1}{\sum m_i^q}U_LM^q U_L^\dag 
= \frac{1}{h_{12}^qh_{23}^q+h_{23}^q+1}\text{diag}(h^q_{12}h^q_{23},h^q_{23},1) 
\label{eq.UMUhie01}
\end{eqnarray}
In the mass hierarchy limit, the leading order mass matrix $M_{0}^q$ can be reconstructed in terms of $U_L^q$ as
\begin{eqnarray}
\frac{1}{{\sum m_i^q}}M_0^q
		=(U_L^q)^\dag\text{diag}(0,0,1)U_L^q
\label{eq.U001U02}
\end{eqnarray}
It is an important implication that only three elements of the matrix, $U^q_{L,13}, U^q_{L,23}$ and $U^q_{L,33}$, have contributions to $M_0^q$.  
Without loss of generality, $U^q_{L,i3}$ can be parameterized by 3 modulus $l_i$ and 3 phases $\theta_i$ as 
		\begin{eqnarray*}
			U^q_{L,33}=l_0e^{i\eta_0},~~~
			\frac{U^q_{L,31}}{U^q_{L,33}}=l_1e^{i\eta_1},~~~
			\frac{U^q_{L,32}}{U^q_{L,33}}=l_2e^{i\eta_2},
		\end{eqnarray*}
(Here $l_i$ and $\eta_i$ shoud be labeled by superscript $^q$. For convenience, we neglect the superscript when there is no possibility of confusion in the following.)
In terms of the unitary of $U_L$,  $l_0$ can be determined by $l_1,l_2$ in terms of
	\begin{eqnarray}
	l_0^2(1+l_1^2+l_2^2)=1.
	\end{eqnarray}
Now,  Eq. (\ref{eq.U001U02}) can be re-written into a factorized form \cite{ZhangCJP2024}
	\begin{eqnarray}
		\frac{1}{\sum m_i^q}M_0^q
		=(K_L^q)^\dag
		M_N^q
		K_L^q.
	\label{eq.KMK2}
	\end{eqnarray}
with diagonal phase matrix $K_L^q=\text{diag}\Big(e^{i\eta_1},e^{i\eta_2},1\Big)$ and real symmetric matrix
\begin{eqnarray}
M_N^q=\frac{1}{l_1^2+l_2^2+1}\Array{ccc}{l_1^2 & l_1l_2 & l_1 \\
			l_1l_2 & l_2^2 & l_2 \\
			l_1 & l_2 & 1
		} .
\label{eq.PatternM}
\end{eqnarray}
Since complex phases in $U_{CKM}$ completely come from family-diagonal $K_L^q$,
it indicates that $K_L^q$ provides a unique origin of CP-violating.
Some rules of mass matrix can be derived from Eq. (\ref{eq.KMK2}).
In the mass hierarchy limit, $M_0^q$ is Hermitian and meets the following rules
	\begin{eqnarray}
		&&\arg[M^q_{0,12}]+\arg[M^q_{0,23}]+\arg[M^q_{0,31}]=0,
		\label{Eq.rules1}\\
		&&\Big|\frac{M^q_{0,11}}{M^q_{0,21}}\Big|=\Big|\frac{M^q_{0,12}}{M^q_{0,22}}\Big|=\Big|\frac{M^q_{0,13}}{M^q_{0,23}}\Big|,
		\label{Eq.rules2}\\
		&&\Big|\frac{M^q_{0,11}}{M^q_{0,31}}\Big|=\Big|\frac{M^q_{0,12}}{M^q_{0,32}}\Big|=\Big|\frac{M^q_{0,13}}{M^q_{0,33}}\Big|.
		\label{Eq.rules3}
	\end{eqnarray}
These rules are valid only if a unitarity condition $U_R^q=U_L^q$ is chosen. 
If releasing the unitarity condition, Eqs. (\ref{Eq.rules1},\ref{Eq.rules2},\ref{Eq.rules3}) may keep their validity for arbitray complex $M_0^q$ as long as replacing $M_0^q$ by $M_0^q(M_0^q)^\dag$. 
Any mass patterns beyond these rules are suppressed by mass hierarchy.

\subsection{$SO(2)^q$ Flavor Symmetry}
Because the phase $K_L^q$ can be absorbed into unitary transformation in the process of diagonalization of $M_0^q$, the mass spectrum of quarks is determined completely by real $M_N^q$. For arbitrary $l_1$ and $l_2$, $M_N^q$ has invariant eigenvalues $(0,0,1)$.  
It is obvious that $M_N^q$ is invariant under a $SO(2)^q$ family rotation $R_N(\theta^q)$ along axis in the direction  of $N=(l_1,l_2,1)$
\begin{eqnarray}
	R_N(\theta^q)M_N^qR_N^T(\theta^q)=M_N^q.
	\label{eq.RMR11}
\end{eqnarray}
Here, the rotation $R_n(\theta)$ along axis $n=(n_x,n_y,n_z)$ can be expressed as
		\begin{eqnarray}
		R_n(\theta)=\begin{pmatrix}
		n_x^2(1-c_\theta)+c_\theta & n_xn_y(1-c_\theta)+n_zs_\theta & n_xn_z(1-c_\theta)-n_ys_\theta \\
			n_xn_y(1-c_\theta)-n_zs_\theta & n_y^2(1-c_\theta)+c_\theta & n_yn_z(1-c_\theta)+n_xs_\theta \\
			n_xn_z(1-c_\theta)+n_ys_\theta & n_yn_z(1-c_\theta)-n_xs_\theta & n_z^2(1-c_\theta)+c_\theta 
		\end{pmatrix}	
	\end{eqnarray}
with $c_\theta=\cos\theta,s_\theta=\sin\theta$.
 
So, a diagonalization transformation of $M_N^q$ can generally be expressed into a product of an orthogonal transformation $S_0$  and an $SO(2)^q$ rotation $R_N(\theta^q)$
\begin{eqnarray}
	&&\Big[ S_0R_N(\theta^q)\Big]M_N^q \Big[S_0R_N(\theta^q)\Big]^T=S_0M_N^qS_0^T=\text{diag}(0,0,1)
	\label{eq.SRMRS}	
\end{eqnarray}
with
\begin{eqnarray}
	S_0= \begin{pmatrix}
	\frac{1}{\sqrt{1+l_1^2}} & 0 & -\frac{l_1}{\sqrt{1+l_1^2}} \\
		-\frac{l_1l_2}{\sqrt{(1+l_1^2)(1+l_1^2+l_2^2)}} & \frac{\sqrt{1+l_1^2}}{\sqrt{1+l_1^2+l_2^2}} & -\frac{l_2}{\sqrt{(1+l_1^2)(1+l_1^2+l_2^2)}} \\
		\frac{l_1}{\sqrt{1+l_1^2+l_2^2}} & \frac{l_2}{\sqrt{1+l_1^2+l_2^2}} & \frac{1}{\sqrt{1+l_1^2+l_2^2}} 
	\end{pmatrix}.	
	\label{eq.generalSexpression}
\end{eqnarray}
For two-fold degenerated eigenvalues of $M_N^q$, the arbitrary choice of $S_0$ comes down to the initial rotation angles $\theta^q$.
The  rotation $R_N(\theta^q)$ in Eq. (\ref{eq.RMR11}) is just $R_3(\theta^q)$ on the basis of mass eigenstates because of 
\begin{eqnarray}
S_0R_N(\theta^q)S_0^T=R_3(\theta^q)=\begin{pmatrix}c_\theta & s_\theta & 0 \\ -s_\theta & c_\theta & 0 \\ 0 & 0 & 1\end{pmatrix}.
\nonumber
\end{eqnarray}
Alternatively, Eq. (\ref{eq.SRMRS}) can also be expressed as
\begin{eqnarray}
\Big[R_3(\theta^q)S_0\Big]M^q_N\Big[R_3(\theta^q)S_0\Big]^T=\text{diag}(0,0,1).
\label{eq.RMRdiag0}
\end{eqnarray}

\subsection{Quark Mixing}
Using Eqs. (\ref{eq.KMK2}) and (\ref{eq.RMRdiag0}), the unitary $U_L^q$ that transforms left-handed quark from gauge eigenstates to mass eigenstates has general form as
\begin{eqnarray}
	U_L^q=R_3(\theta^q)S_0\Big[K_L^q\Big]^\dag
\label{eq.Ulq01}
\end{eqnarray}
So, the CKM mixing matrix is expressed as
\begin{eqnarray}
	U_{CKM}=U_L^u(U_L^d)^\dag=R_3(\theta^u)S_0 ~\text{diag}(e^{i\lambda_1},e^{i\lambda_2},1)S_0^TR_3^T(\theta^d)
\label{eq.Uckm0}
\end{eqnarray}
with the phase difference $\lambda_i$ defined by $\lambda_i=-\eta_i^u+\eta_i^d$.
Only two $\lambda_1$ and $\lambda_2$ rather than all four phases $\eta_{1,2}^{u,d}$ in $K_L^{u,d}$ can contribute to $U_{CKM}$.
$\theta^{u}$ and $\theta^d$ are free parameters in diagonalization transformation of $M_N^q$, however they play a non-trivial role in $U_{CKM}$.  
In the mass limit, quark masses become two-fold degenerate. The transformation $U_L^q$ with any rotation angle $\theta^q$ always diagonalizes quark mass matrix $M_N^q$. 
But in the charged current weak interaction, the $U_{CKM}$ matrix chooses a special set of $\theta^u$ and $\theta^d$. It indicates that  $SO(2)^u\times SO(2)^d$ symmetry is broken in charged current weak interaction.  Eq. (\ref{eq.Uckm0}) shows that the structure of the CKM matrix is determined solely by four free parameters $\theta^u,\theta^d,\lambda_1$ and $\lambda_2$, perfectly corresponding to the three mixing angles and one CP-violating phase observed experimentally, eliminating the complexity that redundant parameterization might bring.

For quarks can be redefined by a free phase in the basis of mass eigenstates, quark mixing can be parameterized by four physical quantities. When adopting a standard form 
\begin{eqnarray}
	U_{CKM}=\begin{pmatrix}1 & 0 & 0 \\ 0 & c_{23} & s_{23} \\ 0 & -s_{23} & c_{23}\end{pmatrix}
		\begin{pmatrix}c_{13} & 0 & s_{13}e^{-i\delta_{CP}} \\ 0 & 1 & 0 \\ -s_{13}e^{i\delta_{CP}} & 0 & c_{13}\end{pmatrix}
		\begin{pmatrix}c_{12} & s_{12} & 0 \\ -s_{12} & c_{12} & 0 \\ 0 & 0 & 1\end{pmatrix}
\end{eqnarray}
with $s_{ij}\equiv \sin\theta_{ij}$ and $c_{ij}\equiv \cos\theta_{ij}$, the mixing angle $\theta_{ij}$ can be determined in terms of $U_{CKM}$ by 
	\begin{eqnarray*}
		s_{13}&=&|U_{CKM,13}|
		\\
		s_{23}^2&=&\frac{|U_{CKM,23}|^2}{1-|U_{CKM,13}|^2}
		\\
		s_{12}^2&=&\frac{|U_{CKM,12}|^2}{1-|U_{CKM,13}|^2}	
	\end{eqnarray*}
and the CP violation phase is determined by Jarlskog's invariant \cite{JarlskogPRL1985}
	\begin{eqnarray}
		J_{CP}=s_{13}c_{13}^2s_{23}c_{23}s_{12}c_{12} \sin\delta_{CP}.
		\label{eq.Jcpmixingangle}
	\end{eqnarray}

\section{The Flat Pattern and Hierarchy Corrections}
\label{sec.correction}
\subsection{The Flat Pattern and $h^1$-order Corrections}
To obtain the physical masses of the first two families and enough precise CKM mixing angles, hierarchy corrections need to be included.
Before hierarchy corrections, two issues need to be considered first: 
	\begin{itemize}
		\item[(1)] the details of the mass pattern, i.e., how to value $l_1,l_2$ in $M_N^q$;
		\item[(2)] how to introduce breakings into $M_N^q$ to generate lighter quark masses.
	\end{itemize}
	
Instead of fitting all possible values of $l_i$ one by one, we care more about a flat pattern with $l_1=l_2=1$. It is because this pattern treats each family indiscriminately and equivalently, any permutation between two families keeps the mass matrix invariant. 
The flat pattern also brings some inspiration to comprehend the principle behind Yukawa interaction.

$M_N^q$ completely determines mass eigenvalues, so any corrections deviated from the rules in Eqs. (\ref{Eq.rules2},\ref{Eq.rules3}) will generates a correction eigenvalues departuring from $(0,0,1)$. 
Since diagonal elements of mass matrix can be used to define parameters $l_1$ and $l_2$,
the correction of normalized mass matrix manifests as real symmetric form with three real perturbations $\delta_{12},\delta_{23},\delta_{13}$ in non-diagonal elements.
Thus, the flat pattern with hierarchy correction becomes
	\begin{eqnarray}
		{M}_\delta^q=\frac{1}{3}\begin{pmatrix}
		1 & 1+\delta_{12}^q & 1+\delta_{13}^q\\ 1+\delta_{12}^q & 1 & 1+\delta_{23}^q \\ 1+\delta_{13}^q & 1+\delta_{23}^q & 1
		\end{pmatrix}
		\label{eq.MdeltaQ0}
	\end{eqnarray}
Here, the subscript $_\delta$ labels hierarchy corrections.

Compared with the democratic matrix that is often used for neutrinos and also quarks \cite{SogamiPTP1998,FritzschCPC2017}, 
the flat pattern is proposed based on phenomenological features rather than model assumptions. There are some obvious differences between two kinds of patterns
\begin{itemize}
	\item[(1)] complex phases are completely factorized into $K_L^q$ in terms of the factorized structure of mass matrix in Eq. (\ref{eq.KMK2}) in the flat pattern. however, there is no guidance in the democratic matrix;
	\item[(2)] the hierarchy corrections are expressed by 3 real $\delta^q_{ij}$ in non-diagonal elements of mass matrix in the flat pattern, rather than some random complex corrections in democratic matrix;
	\item[(3)] the flat pattern is used to up-type and down-type quarks at the same time because of their common hierarchy. The democratic matrix is often used to down-type quarks in the diagonal up-type mass basis.
\end{itemize}

In the first order of $h_{ij}^q$, to obtain the eigenvalues $(0, h_{23}^q,1-h_{23}^q)$ of normalized $M_\delta^q$,  the solutions of $\delta_{ij}^q$ are given by 
		\begin{eqnarray}
			\delta_{12}^q&=&\left(-\frac{3}{4}\cos(2\theta^q)-\frac{9}{4\sqrt{3}}\sin(2\theta^q)-\frac{3}{2}\right)h_{23}^q,~~
			\label{Eq.Rotationthetah1}\\
			\delta_{23}^q&=&\left(-\frac{3}{4}\cos(2\theta^q)+\frac{9}{4\sqrt{3}}\sin(2\theta^q)-\frac{3}{2}\right)h_{23}^q,~~
			\label{Eq.Rotationthetah2}\\
			\delta_{13}^q&=&\left(\frac{3}{2}\cos(2\theta^q)-\frac{3}{2}\right)h_{23}^q
			\label{Eq.Rotationthetah3}
		\end{eqnarray}
The solutions indicate that the close-to-flat mass matrix $M_\delta^q$ keeps the $SO(2)^q$ rotation symmetry in the order of $h^1$.
Defined $M^q_\delta(\theta^q)$ as
		\begin{eqnarray}
			M^q_\delta(\theta^q)=\frac{1}{3}\begin{pmatrix}1 & 1+\delta_{12}(\theta^q) & 1+\delta_{13}(\theta^q)  \\
					1+\delta_{12}(\theta^q) & 1 & 1+\delta_{23}(\theta^q) \\
					1+\delta_{13}(\theta^q)  & 1+\delta_{23}(\theta^q) & 1\end{pmatrix},
		\end{eqnarray}
	it can be expressed in terms of rotation $R_\delta(\theta)$ as
	\begin{eqnarray}
		M^q_\delta(\theta^q)=R_{\delta}^T(\theta^q)M^q_\delta(0)R_{\delta}(\theta^q).
	\label{Eq.MthetaM0}
	\end{eqnarray}
Here, $M^q_\delta(0)$ is the mass matrix corresponding to the initial values 
	$$(\delta_{12}^q,\delta_{23}^q,\delta_{13}^q)=-\frac{9}{4}h_{23}^q(1,1,0),$$
	and $R_\delta(\theta)$ is a $SO(2)^q$ rotation along the corrected direction $\left(1,1-\frac{9}{4}h_{23}^q,1\right)$.
		
		Defined an orthogonal transformation $S_\delta^q$ meeting
		\begin{eqnarray}
		S_\delta^qM_\delta^q(0)(S_\delta^q)^T
		=\text{diag}(0,h_{23}^q,1-h_{23}^q),
		\end{eqnarray}
		$S_\delta^q$ can be expanded as a power series of $h$ 
		\begin{eqnarray}
			S_\delta^q=S_0+S^q_{h1}+\mathcal{O}(h^2).
		\end{eqnarray}
After some calculation, leading order $S_0$ and 1-order correction $S_{h1}$ are
		\begin{eqnarray}
			S_0&=&\begin{pmatrix}\frac{1}{\sqrt{2}} & 0 &-\frac{1}{\sqrt{2}} \\
			-\frac{1}{\sqrt{6}} & \sqrt{\frac{2}{3}} & -\frac{1}{\sqrt{6}}\\
	\frac{1}{\sqrt{3}} & \frac{1}{\sqrt{3}} & \frac{1}{\sqrt{3}} \end{pmatrix}
			\\
			S^q_{h1}&=&\frac{1}{4\sqrt{3}}h_{23}^q\begin{pmatrix}0 & 0 &0 \\ \sqrt{2} & \sqrt{2} & \sqrt{2} \\ 1 & -2 &1\end{pmatrix}
		\end{eqnarray}
Generally, the $M_\delta^q(\theta^q)$ can be diagonalized by $S_\delta^qR_\delta(\theta)$
	\begin{eqnarray}
		\Big[S_\delta^qR_\delta(\theta^q)\Big]M_\delta^q(\theta^q)\Big[S_\delta^qR_\delta(\theta^q)\Big]^T
		=\text{diag}(0,h_{23}^q,1-h_{23}^q)+\mathcal{O}(h^2).
	\label{Eq.MdeltaDiag}
	\end{eqnarray}
The CKM matrix with the 1-order hierarchy correction becomes
	\begin{eqnarray}
		U_{CKM}&=&\Big[S_\delta^uR_{\delta}(\theta^u)\Big]\Big[\text{diag}(1, e^{i\lambda_1^u},e^{i\lambda_2^u})\Big]\Big[R_{\delta}^T(\theta^d)(S_\delta^d)^T\Big].
		\label{Eq.ClostToFlatCKM01}
	\end{eqnarray}
\subsection{$h^2$-order Corrections}
Because of $m_1^q/\sum m_i^q= h_{23}^qh_{12}^q\sim \mathcal{O}(h^2)$, to address the lightest quark, the $h^2$ order corrections need  to be considered.  To generate three quark masses up to the leading order
		\begin{eqnarray}
			\frac{m_1}{\sum m_i^q}\simeq h^q_{12}h^q_{23},~
			\frac{m_2}{\sum m_i^q}\simeq h^q_{23},~
			\frac{m_3}{\sum m_i^q}\simeq 1-h^q_{23},
		\end{eqnarray}
		the general solutions of $\delta_{ij}^q$ can be soloved perturbatively as 
		\begin{eqnarray}
			\delta_{12}^q&=&\left(-\frac{3}{4}\cos(2\theta^q)-\frac{9}{4\sqrt{3}}\sin(2\theta^q)-\frac{3}{2}\right)h_{23}^q
				-3h^q_{12}h^q_{23}
				+\frac{3}{16}(h^q_{23})^2(\cos(6\theta^q)-1)+\mathcal{O}(h^3),
			\label{eq.deltarotationh21}
				\\
			\delta_{23}^q&=&\left(-\frac{3}{4}\cos(2\theta^q)+\frac{9}{4\sqrt{3}}\sin(2\theta^q)-\frac{3}{2}\right)h_{23}^q
				-3h^q_{12}h^q_{23}
				+\frac{3}{16}(h^q_{23})^2(\cos(6\theta^q)-1)+\mathcal{O}(h^3),
			\label{eq.deltarotationh22}
				\\
			\delta_{13}^q&=&\left(\frac{3}{2}\cos(2\theta^q)-\frac{3}{2}\right)h_{23}^q
					-3h^q_{12}h^q_{23}
					+\frac{3}{16}(h^q_{23})^2(\cos(6\theta^q)-1)+\mathcal{O}(h^3)
			\label{eq.deltarotationh23}
		\end{eqnarray}
By the same way, we find that $M_\delta^q$ remains an approximate $SO(2)^q$ family symmetry at the order of $h^2$. 
$U_{CKM}$ keeps the same factorized structure as one in hierarchy limit.

\subsection{CP Violation as a Check Point}
Unlike the CKM mixing angles with a value of close to zero, the CP-violating phase is far from zero. 
It hints that $\delta_{CP}$ can not arise from mass hierarchy corrections. For this reason, the origin of CP violation provides a checkpoint of flavor physics models. A successful flavor model must be able to give a mechanism to yield a non-vanishing CP violation in the mass hierarchy limit. 

From Eq. (\ref{eq.Uckm0}), the Jarlskog invariant can be obtained as follows:
	\begin{eqnarray}
		J_{CP}&=&\frac{1}{54}\Bigg\{\sqrt{3}\sin(\frac{\lambda_2}{2})\sin(2\theta^u+2\theta^d)\Big[\cos(\lambda_1-\frac{3\lambda_2}{2})+\cos(\lambda_1+\frac{\lambda_2}{2})-2\cos(\frac{\lambda_2}{2})\Big]
		\nonumber\\
		&&+\sin(\lambda_1-\frac{\lambda_2}{2})
			\Big[
			-4\cos(\lambda_1-\frac{\lambda_2}{2})\sin(2\theta^u)\sin(2\theta^d)
			\nonumber\\
			&&+\cos(\frac{3\lambda_2}{2})[2\cos(2\theta^u-2\theta^d)+\cos(2\theta^u+2\theta^d)]
			+3\sin(\frac{\lambda_2}{2})\sin(2\theta^u+2\theta^d)\Big]
		\Bigg\}
\		\label{Eq.Jcp0}
	\end{eqnarray}
The expression reveals the mechanism of CP violation: when the phase $\lambda_1,\lambda_2\neq 0$, CP violation can still occur through the interference effect of the rotation angles $\theta^u$ and $\theta^d$ even in the limit of mass degeneracy for the first two generations of quarks. This indicates that the observed large CP phase is not due to mass hierarchy corrections but is a direct consequence of $SO(2)^{q}$ family symmetry breaking.

\section{Fit Results}
\label{sec.fit}
The flat pattern provides a minimal parameterization of flavor structure. The total family mass of quarks is determined by family universal Yukawa coupling $y^{u,d}$.
In each type quark, three $\delta_{ij}^q$ determine two hierarchies $h_{12}^q$ and $h_{23}^q$ and left degree of freedom is $SO(2)^q$ rotation angle $\theta^{q}$.
The $\theta^q$  enters the mixing matrix in charged current weak interaction and parameterizes $U_{CKM}$ along with complete phases $\lambda_{1,2}$ \cite{ZhangJPG2023}.

 To fit to quark masses and the CKM mixing, we can initially $\delta_{ij}^q$ in terms of  Eqs. (\ref{eq.deltarotationh21},\ref{eq.deltarotationh22},\ref{eq.deltarotationh23}) with $\theta=0$, 
 and then scan four parameters, $\theta^{u},\theta^d,\lambda_{1}$ and $\lambda_2$, in the whole rangle to fit the CKM mixing.

 A best fit point is obtained at
	\begin{eqnarray}
		\theta^u=0.5013~
		,\theta^d=3.416~
		,\lambda_1=0.04709~
		,\lambda_2=6.228.
	\end{eqnarray}
And the fit results are listed in Tab. \ref{tab.fit00}. 
\begin{table}[htp]
\caption{CKM mixing Fit. The experiment data are from ref. \cite{PDG2024} .}
\begin{center}
\begin{tabular}{|c|c|c|}
\hline
\hline
 para. & exp. & fit
 \\
 \hline
 $s_{12}$ & $0.22501\pm0.00068$ & $0.2248$
 \\
 $s_{23}$ & $0.0413^{+0.00079}_{-0.00069}$ & $0.04179$
 \\
 $s_{13}$ & $0.003773^{+0.000090}_{-0.000085}$ & $0.003750$
 \\
 $\delta_{CP}$ & $1.147\pm0.026$ & $1.141$
 \\
 \hline
\hline
\end{tabular}
\end{center}
\label{tab.fit00}
\end{table}%
At the fit point, the flat mass matrixes are decomposed into leading order and hierarchy corrections as
	\begin{eqnarray}
	M_N^u&=&\frac{1}{3}\begin{pmatrix}
		1 & 1 & 1 \\ 1 & 1 & 1 \\ 1 & 1 & 1
		\end{pmatrix}
		+\begin{pmatrix}
		-0.00001753 & -0.007355 & -0.001734 \\
		-0.007355 & 0.00004075 & -0.001982 
		\\
		-0.001734 & -0.001982 & -0.00001074
		\end{pmatrix}
	\\
	M_N^d&=&\frac{1}{3}\begin{pmatrix}1 & 1 & 1 \\ 1 & 1 & 1 \\ 1 & 1 & 1\end{pmatrix}
		+\begin{pmatrix}-0.001127 & -0.02047 & -0.003732 
		\\
		-0.02048 & 0.003043 & -0.01053 
		\\
		-0.003732 & -0.01053 & -0.0008045\end{pmatrix}
	\end{eqnarray}
The mass hierarchies are shown in Tab. \ref{tab.hierarchy}
\begin{table}[htp]
\caption{CKM mixing Fit.}
\begin{center}
\begin{tabular}{|c|c|c|}
\hline
\hline
 para. & exp. & fit
 \\
 \hline
 $h_{12}^u$ & $0.001697\pm0.000055$ & $0.001697$
 \\
 $h_{23}^u$ & $0.007370\pm0.000029$ & $0.007425$
 \\
 $h_{12}^u$ & $0.05027\pm0.00084$ & $0.05027$
 \\
 $h_{23}^u$ & $0.02272\pm0.00020$ & $0.02326$
\\
 \hline
\hline
\end{tabular}
\end{center}
\label{tab.hierarchy}
\end{table}%

These fit results show that the CKM mixing parameters are dominated by angles $\theta^{u,d}$ and phases $\lambda_{1,2}$. 
And hierarchy only contributes small corrections. 
In particular, the primary part of mass matrix stems from the flat pattern, which reasonably explains the large value of CP-violating phase in the mass hierarchy limit. All results support the effectiveness of the flat pattern in describing quark masses and mixing.

\section{Summary}
\label{sec.summary}
 Inspired by the observed quark mass hierarchy, a factorized mass matrix has been derived, and a set of rules for mass matrices has been proposed to systematically examine the flavor structure in models. These model-independent rules provide a clear framework, allowing us to move beyond uncertainty when exploring flavor structures and deciphering the nature of Yukawa interactions. As a universal family structure, the flat pattern is analyzed in detail up to $\mathcal{O}(h^2)$, demonstrating its ability to accurately reproduce quark masses and mixing parameters.

Although the flat pattern originates from the quark mass hierarchy, it can be generalized to the lepton sector as well. In addition to charged leptons, neutrino masses also exhibit a hierarchical structure—either 
$m_1^\nu<m_2^\nu \ll m_3^\nu$ for the normal ordering or $m_3^\nu\ll m_1^\nu<m_2^\nu$ for the inverted ordering. This generalization may pave the way for constructing a universal mass pattern in future studies.

\section*{Acknowledgments}
This work is supported by Shaanxi Foundation SAFS 22JSY035 of China.


\end{document}